\documentclass[aps,prd,a4paper,preprintnumbers,amsmath,amssymb,superscriptaddress,twocolumn,floatfix,showpacs,showkeys,nofootinbib]{revtex4}
 \usepackage{graphics}

\begin{document}

\title{Cosmology of Mass-Varying Neutrinos Driven by Quintessence: Theory and Observations} 
\author{A.~W.~Brookfield} 
\affiliation{Department of
Applied Mathematics and Department of Physics, Astro--Particle Theory
$\&$ Cosmology Group, Hounsfield Road, Hicks Building, University of
Sheffield, Sheffield S3 7RH, United Kingdom} 
\author{C.~van de Bruck}
\affiliation{Department of Applied Mathematics, Astro--Particle Theory
$\&$ Cosmology Group, Hounsfield Road, Hicks Building, University of
Sheffield, Sheffield S3 7RH, United Kingdom} 
\author{D.~F.~Mota}
\affiliation{Institute of Theoretical Astrophysics, University of
Oslo, 0315 Oslo, Norway} 
\author{D.~Tocchini-Valentini}
\affiliation{Astrophysics Department, Oxford University, Keble Road,
Oxford OX1 3RH, United Kingdom\\
Department of Physics and Astronomy, The Johns Hopkins University, Baltimore, MD 21218, USA}

\date{14 December 2005}

\begin{abstract}
The effects of mass-varying neutrinos on cosmic microwave background
(CMB) anisotropies and large scale structures (LSS) are studied. In
these models, dark energy and neutrinos are coupled such that the
neutrino masses are functions of the scalar field playing the role of
dark energy. We begin by describing the cosmological background
evolution of such a system. It is pointed out that, similar to models
with a dark matter/dark energy interaction, the apparent equation of
state measured with SNIa can be smaller than -1. We then discuss the
effect of mass-varying neutrinos on the CMB anisotropies and the
matter power spectrum. A suppression of power in the CMB power
spectrum at large angular scales is usually observed. We give an
explanation for this behaviour and discuss different couplings and
quintessence potentials to show the generality of the results
obtained.  We perform a likelihood analysis using wide-ranging SNIa,
CMB and LSS observations to assess whether such theories are viable.
Treating the neutrino mass as a free parameter we find that the 
constraints on the coupling are weak, since CMB and LSS surveys give only upper bounds 
on the neutrino mass. However, fixing a priori the neutrino masses, we find that there is 
some evidence that the existence of such a coupling is actually preferred by current
cosmological data over the standard $\Lambda$CDM cosmology. 
\end{abstract}


\maketitle

\section{Introduction}

Recent cosmological observations indicate that the expansion of the 
universe is accelerating \cite{observation1}-\cite{observation3}. 
It follows from General Relativity that the
dominant energy component today must have negative pressure. Many
candidates have been proposed over the years, including scalar field
models, which are well motivated from the point of view of particle
physics theories, see e.g. \cite{wetterich}-\cite{wetterich2}. The
main prediction of these types of models is that the dark energy
equation of state becomes a dynamical quantity, and can vary from the
usual value of $w=-1$ for a cosmological constant. Although such
models are very attractive, they are plagued with several theoretical
difficulties, such as the stability of the potential under quantum
corrections \cite{doran} or why the dark energy scalar field seems not
to mediate a force between normal matter particles \cite{carroll}.  In
addition, the energy scale of the scalar field is put in by hand and
usually not connected to a more fundamental energy scale. However,
attempts have been made to address these problems, such as models with
ultralight pseudo-Nambu-Goldstone bosons (see, for example,
\cite{frieman}, \cite{nilles} and \cite{kaloperdark}; for a review,
see \cite{pecceireview}).

It is expected that any explanation for dark energy will involve
physics beyond the standard model of particle physics. Recently, a new
class of models have been proposed, which entertain the idea of a
possible connection between neutrinos and dark energy.  Their
theoretical and observational consequences have already been studied
very extensively \cite{neutrinobeg}-\cite{neutrinoend}. The main
motivation for a connection between dark energy and neutrinos is that
the energy scale of dark energy (${\cal O}(10^{-3})$~eV) is of the
order of the neutrino mass scale. In these models the neutrino mass
scale and the dark energy scale are linked to each other, and hence
the observed non-zero neutrino masses (see
\cite{massiveneutrinos1}-\cite{massiveneutrinos3}) cannot be
understood without an understanding of dark energy. Also, one may hope
that these models might provide an explanation for the coincidence
problem \cite{fardon1}.

In this paper we investigate the cosmology of neutrino models of dark
energy.  We take into account the {\it full} evolution of the
neutrinos, i.e. studying the relativistic and non-relativistic regimes
and the transition in between.

Armed with a complete numerical model for the evolution of the coupled
neutrinos, we compare the background evolution with Supernova data and
study how the modified perturbations affect the cosmic microwave
background radiation (CMB) temperature anisotropies 
and large scale structures (LSS) matter power spectrum. We thereby
present the details of the results outlined in \cite{us} and discuss
other forms of coupling between dark energy and neutrinos and
potentials for the dark energy field.

The paper is organized as follows: In Section II we discuss the
background evolution of the coupled dark energy-neutrino system in the
context of a typical quintessential potential.  In Section III we derive
the evolution equations for cosmological perturbations in neutrino
models of dark energy, and present the modified CMB and matter power
spectra. In Section IV we discuss other couplings and potentials, such
as inverse power-law potentials and field--dependent couplings. In
Section V we compare our theory with data, using a public Markov-Chain
Monte-Carlo data analysis program.  We conclude in Section VI.

\section{The Cosmological Background Evolution}

In a flat, homogeneous, Friedmann--Robertson--Walker universe with
line-element
\begin{equation}
ds^2 = a^2(\tau)\left(-d\tau^2 + \delta_{ij}dx^i dx^j \right),
\end{equation}
the Einstein equations describe the evolution of the scale factor
$a(\tau)$:
\begin{eqnarray}
H^2\equiv\left(\frac{\dot{a}}{a}\right)^2 &=& \frac{8\pi}{3}Ga^2\rho,\\
\frac{d}{d\tau}\left(\frac{\dot{a}}{a}\right)&=&-\frac{4\pi}{3}Ga^2(\rho+3p).
\end{eqnarray}
In these equations, $\rho(\tau)$ and $p(\tau)$ are the total energy
density and pressure respectively and the dot refers to the derivative
with respect to conformal time $\tau$.  Defining
$\Omega_i=\rho_i/\rho_c$, where $\rho_c$ is the critical energy
density for a flat universe and $\rho_i$ are the energy densities of
the individual matter species, the equations above require that
$\Omega=\sum_i \Omega_i =1$. In the following we will set $8\pi G \equiv 1$. 

In our model we consider a universe with the usual energy--matter
composition. At early times, the energy density is dominated by the
relativistic species -- radiation and highly relativistic neutrinos.
As the universe expands the energy density in radiation decays, and
the universe becomes matter dominated.  The dominant matter species is
assumed to be Cold Dark Matter (CDM), which is non--relativistic,
weakly interacting and behaves like a perfect pressureless fluid.  At
this time there are also contributions to $\Omega$ from baryons and
neutrinos (which having cooled behave in a manner similar to CDM).  At
late times the matter energy densities also decay away, and we enter
the dark energy dominated epoch. In common with standard quintessence
models we describe the dark energy sector using a dynamical scalar
field with potential $V(\phi)$, where the form of the potential is
chosen (and fine--tuned) to produce the necessary late time
acceleration. The energy density and pressure of the scalar field are
defined by the usual expressions,

\begin{eqnarray}
\rho_\phi&=&\frac{1}{2a^2}\dot{\phi}^2+V(\phi) \label{rhop}\\
p_\phi&=&\frac{1}{2a^2}\dot{\phi}^2-V(\phi). \label{pp}
\end{eqnarray}

In this paper, we consider the consequences of a coupling between
neutrinos and dark energy. To describe this coupling, we follow
\cite{fardon1}: the coupling of dark energy to the neutrinos results
in the neutrino mass becoming a function of the scalar field,
i.e. $m_\nu=m_\nu(\phi)$, and so the mass of the neutrinos changes as
the scalar field evolves. For our purposes it does not matter if the
neutrinos are Majorana or Dirac particles, and for simplicity
we assume three species of neutrinos with degenerate mass\footnote{In fact, such 
an assumption is quite natural and has no strong consequences to this work. 
In the mass regions detectable in astronomical observations, the three 
neutrino masses are nearly degenerate. Adding to that, cosmology is in leading order sensitive to $\sum m_{\nu}$.}
It is well known \cite{doran} that the light mass of the quintessence potential
results in it being highly unstable to radiative corrections, and that
the addition of a coupling between the dark energy and other matter
species only serves to further exacerbate this problem.  In this
regard it is important that both the quintessence potential and the
neutrino mass are regarded as classical, effective quantities, which
already include radiative corrections.

It is also important to note that our theory differs significantly in
one key aspect from the work of \cite{fardon1}.  In our models, the
dark energy sector is described by a \emph{light} scalar field, with a
mass which is at most of order $H$.  The potential chosen by Fardon et
al. was such that the mass of the scalar field is much larger than $H$
for most of its history, and this can have significant implications
upon the behaviour of the neutrino background and the growth of
perturbations \cite{zalda} as we will discuss later.

To fully describe the evolution of cosmological neutrinos, we must
calculate their distribution function $f\left(x^i,p^i,\tau\right)$ in
phase space.  An important fact to note is that even though the
neutrinos interact with dark energy, we treat the interaction
classically and, as will be shown in eq. (\ref{eq:action}), they can
be thought as free-falling in a metric given by
\begin{equation}
g_{\alpha \beta}^{\nu} = m_{\nu}(\phi)^{2}g_{\alpha \beta}.
\end{equation}
Thus, the theory we are going to consider is a special type of scalar-tensor theory, 
in which the scalar degree of freedom couples only to neutrinos. 
It follows that the neutrino phase-space density is incompressible and
we can treat the neutrinos as collisionless particles throughout the
period of interest as long as we keep track of the evolution of the
neutrino mass. We therefore need to solve the Boltzmann equation in
collisionless form simultaneously with the scalar field evolution
equations. Once the distribution function is known, the pressure and
energy density of the neutrinos can be calculated. In this Section we
will discuss the background evolution only; in the next Section we
will discuss cosmological perturbations in these models.

The energy density stored in the neutrinos is given by
\begin{equation}\label{eq:density}
\rho_\nu=\frac{1}{a^4}\int q^2 dq\, d\Omega \epsilon f_0(q),
\end{equation}
and the pressure by
\begin{equation}\label{eq:pressure}
p_\nu=\frac{1}{3a^4}\int q^2 dq\, d\Omega f_0(q) \frac{q^2}{\epsilon},
\end{equation}
where $f_0(q)$ is the usual unperturbed background neutrino
Fermi-Dirac distribution function
\begin{equation}
f_0(\epsilon)=\frac{g_s}{h_P^3}\frac{1}{e^{\epsilon/k_B T_0}+1},
\end{equation}
and $\epsilon^2 = q^2 + m_\nu^2(\phi)a^2$ ($q$ denotes the comoving
momentum).  As usual, $g_s$, $h_P$ and $k_B$ stand for the number of
spin degrees of freedom, Planck's constant and Boltzmann's constant
respectively. In the following we will assume that the neutrinos
decouple whilst they are still relativistic, and therefore the
phase-space density only depends upon the comoving momentum. Taking
the time-derivative of eq.~(\ref{eq:density}), it can be easily shown
that
\begin{equation}\label{eq:nuenergy}
\dot{\rho}_\nu +3H\left(\rho_\nu+p_\nu\right) = \frac{d \ln m_\nu}{d
\phi} \dot{\phi} \left(\rho_\nu - 3 p_\nu\right).
\end{equation}

We describe the dark energy sector using a scalar field with potential
$V(\phi)$.  Taking into account the energy conservation of the coupled
neutrino--dark energy system, one can immediately find that the
evolution of the scalar field is described by a modified Klein-Gordon
equation
\begin{equation}\label{eq:kleingordon}
\ddot{\phi}+2H\dot{\phi}+a^2\frac{dV}{d\phi}=-a^2\frac{d \ln m_\nu}{d
\phi}\left(\rho_\nu-3p_\nu\right).
\end{equation}
This equation contains an extra source term with respect to the
uncoupled case, which accounts for the energy exchange between the
neutrinos and the scalar field.

For the remainder of this Section and the next, we consider a typical
exponential form for the dark energy potential, namely
\begin{equation}
V(\phi)=V_0 e^{-\sigma \phi}
\end{equation}
and define $\sigma \equiv\sqrt{\frac{3}{2}}\lambda$.  We also choose
to take
\begin{equation}
m_\nu(\phi)=M_0e^{\beta \phi}
\end{equation}

The exponential potential can produce a non-scaling solution that may give late time acceleration, 
depending upon the steepness
of the potential, $\sigma$ (see e.g. \cite{copeland,ferreira,amendola1,
domenico1,kallosh}).  In an uncoupled system with 
$\sigma<\sqrt{6}$ 
there 
exists a critical point that is stable for $\sigma^2<3(1+w)$, where $w$ stands for the equation of state of matter or radiation, and in which $\Omega_\phi=1$.
This solution will lead to 
acceleration provided that $\sigma<\sqrt{2}$. 
The existence of scaling solutions depends upon the equation of state of the other components present in the universe.
Choosing $\sigma^2>3(1+w)$ leads to a scaling solution with 
$\Omega_\phi=3(1+w)/\sigma^2$ \cite{copeland}. (See also \cite{weller2}, who use 
the exponential potential as a dark energy model.)  The requirement that the present 
day dark energy density is $\Omega_\phi \sim 0.7$ is hard to reconcile with the
scaling solution at early times, since 
in this case it follows that $\Omega_\phi=4/\sigma^2$, 
whilst
big bang nucleosynthesis requires that the dark energy density in the early universe 
is very small \cite{bean}. 

In this Section we focus our discussion on models with $\sigma<\sqrt{2}$, which with 
an appropriate choice of $V_0$, can provide late-time acceleration with 
$\Omega_\phi \sim 0.7$ today (note that the late-time attractor 
$\Omega_\phi=1$ lies in the future).  This 
choice of $\sigma$ also ensures that the energy density in the form of dark 
energy at the time of BBN is very small, because for early times the 
quintessence field is frozen and acts like a cosmological constant with 
an energy density similar to the observed dark energy density today.  

The presence of neutrino coupling can 
potentially
affect this result, as the 
coupled field begins to evolve at earlier redshifts ($z \sim 10^7$), however 
the fraction of the total energy stored in the scalar field at early times
remains insignificant.  Note that this choice of potential reduces to the 
cosmological constant case for a perfectly flat potential with zero coupling. 

From the neutrino energy conservation equation (\ref{eq:nuenergy}),
and for our choice of $m_\nu(\phi)$ and $V(\phi)$, one can see that
the dynamics of the scalar field can be described by an {\it effective
potential}
\begin{equation}
V_{\rm eff}=V(\phi) +
\left(\tilde{\rho}_\nu-3\tilde{p}_\nu\right)e^{\beta \phi},
\end{equation}
where $\tilde{\rho}_\nu \equiv \rho_\nu e^{-\beta \phi}$ and
$\tilde{p}_\nu \equiv p_\nu e^{-\beta \phi}$ are independent of $\phi$. 
It can be shown that the effective potential will only have a minimum
when $\beta \sigma>0$. For the neutrinos we have numerically evaluated
the integrals (\ref{eq:density}) and (\ref{eq:pressure}), which then
have been used in the Klein--Gordon eq.  (\ref{eq:kleingordon}) to
find the evolution of the scalar field.

Figure 1 shows some typical examples of how the coupling of the
neutrinos to the scalar field causes the mass of the neutrinos to
evolve with time.  Deep within the radiation dominated epoch, at times
when the neutrinos are highly relativistic, the scalar field is Hubble
damped and therefore the neutrino mass is (almost) constant.
For quintessence models, $\dot \phi$ is at most of order $H$, and
therefore for relativistic species the coupling term in equation
(\ref{eq:nuenergy}) is clearly suppressed relative to the Hubble
damping term.  As the universe expands the neutrinos cool and become
non--relativistic at a temperature corresponding to the neutrino mass.
Hence, the extra coupling terms in equations (\ref{eq:nuenergy}) and
(\ref{eq:kleingordon}) become more and more important, allowing energy
to be exchanged between the neutrinos and the scalar field. This
interaction causes the scalar field, and hence the neutrino mass, to
evolve.

It is important to note the two different types of behaviour seen in
Figure 1 for the evolution of the neutrino mass. For models which have
$\beta \sigma>0$ the effective potential possesses a minimum, and
after some time the field passes through this minimum, slows down,
stops and eventually rolls back towards the minimum. For models which
do not posses an effective minimum $\dot{\phi}$ is always negative,
and the scalar field will continue to roll down the effective
potential unimpeded.  We compare the behaviour of our light scalar
field with the heavy acceleron field used in \cite{fardon1} - in their
model the scalar field sits in the effective minimum of its potential
for most of the time during the cosmic history, and it is the
evolution of the effective minimum which drives the dynamics of the
neutrino mass.  As discussed in \cite{fardon1}, the mass of their
neutrinos increases as the universe expands, whereas in our model the
neutrinos are heavier in the past and become lighter
(although as we will discuss in Section IV, suitable choices of
coupling and potential can realize coupled neutrino--quintessence
models with neutrinos which are lighter in the past).

\begin{figure}
\centerline{\scalebox{0.5}{\includegraphics{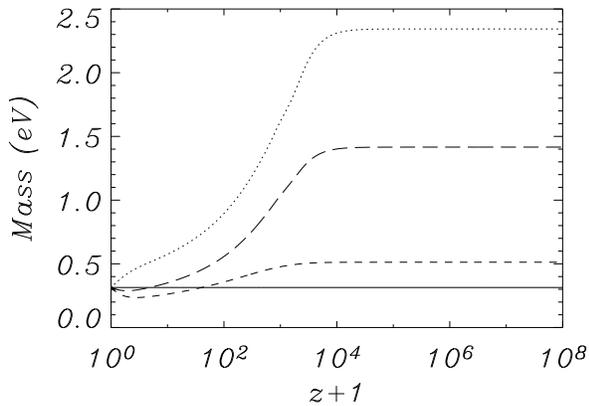}}}
\caption{Plot showing the evolution of the neutrino mass for an
exponential potential and coupling (solid line: $\beta=0$,
$\lambda=1$; short dashed line: $\beta=1$, $\lambda = 1$; dotted line:
$\beta=-0.79$, $\lambda = 1$; long dashed line $\beta = 1$, $\lambda =
0.5$.). In all models, we have arranged that $m_{\nu} = 0.314$~eV
today.}
\end{figure}
\begin{figure}
\centerline{\scalebox{0.4}{\includegraphics{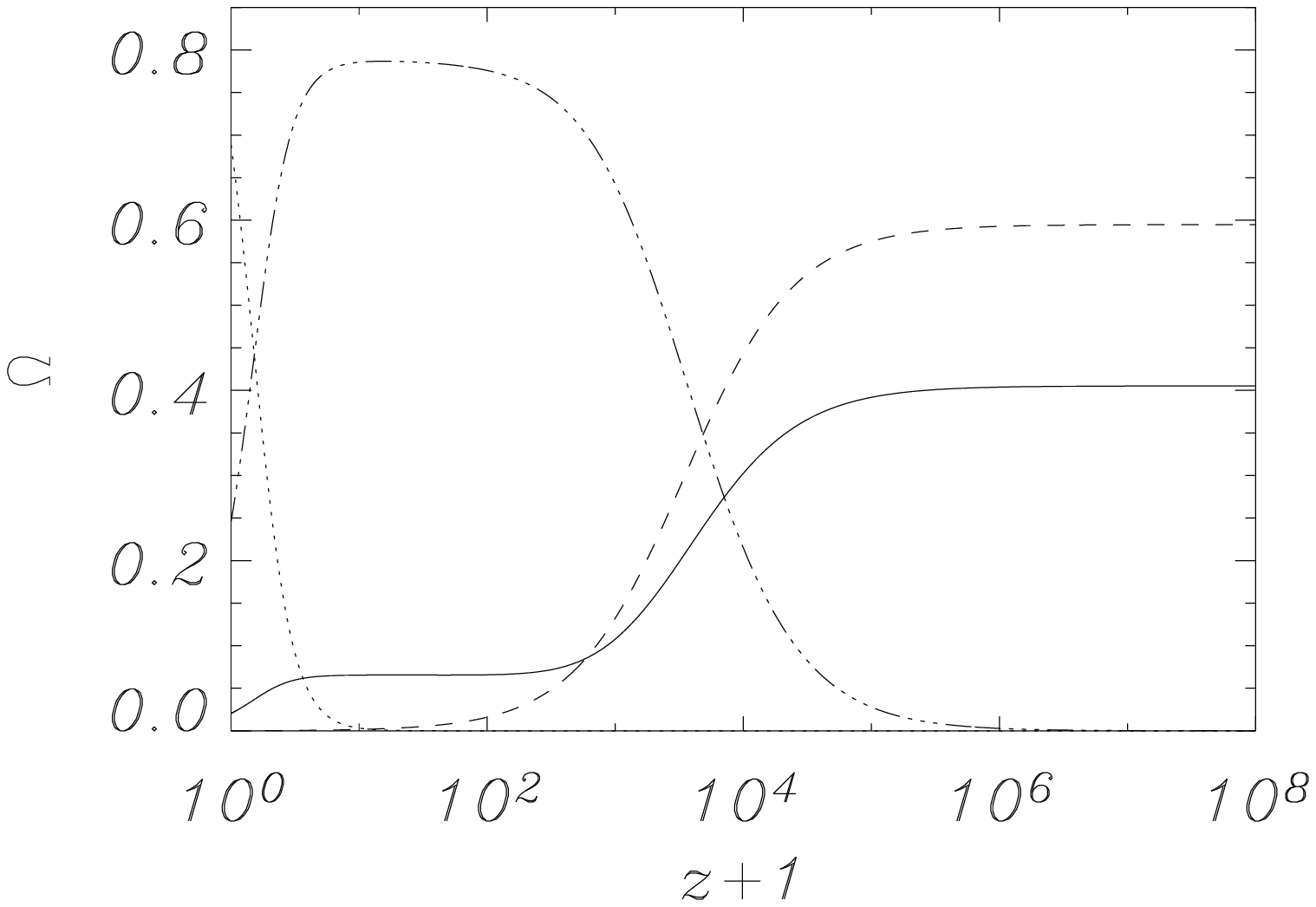}}}
\centerline{\scalebox{0.4}{\includegraphics{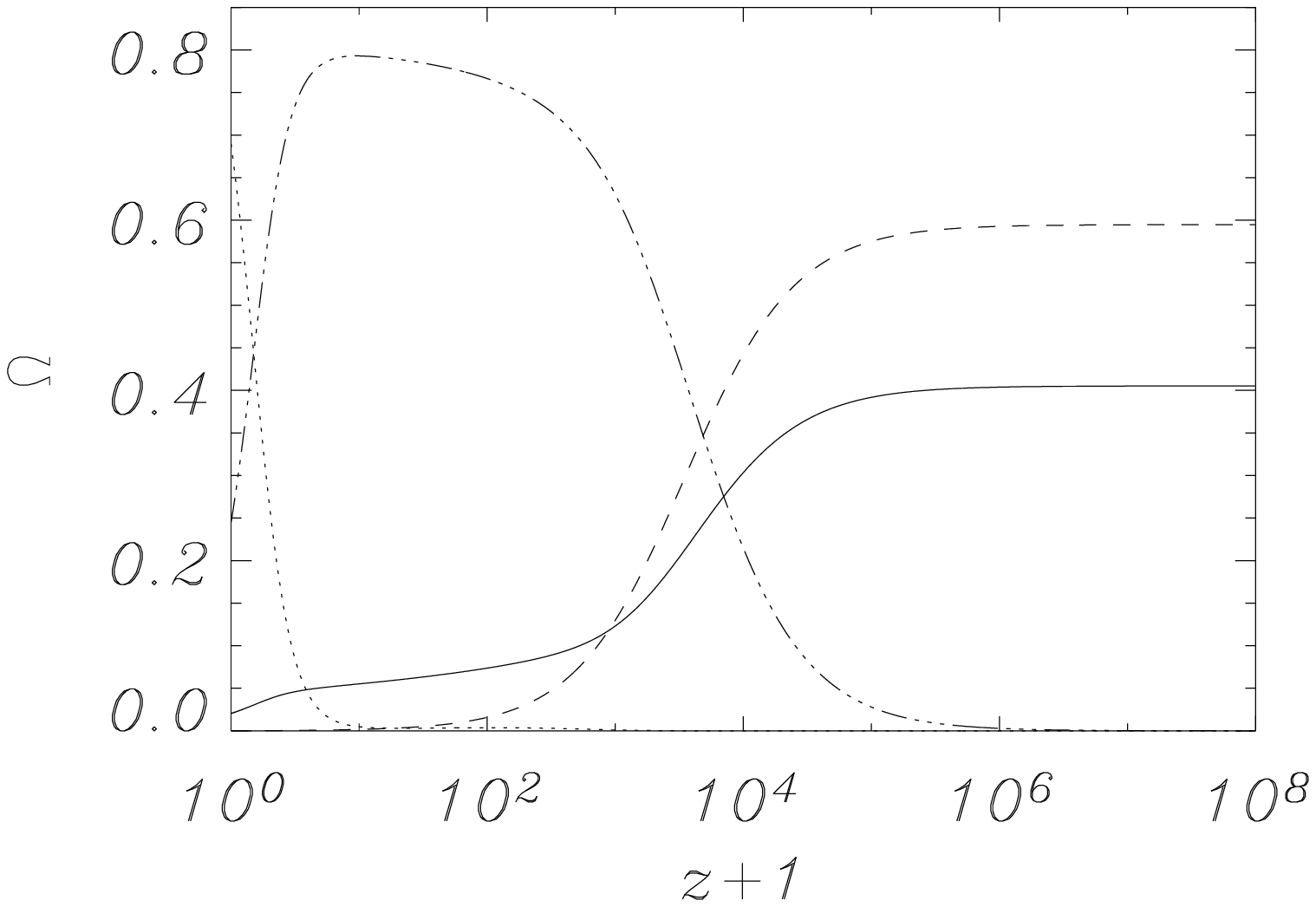}}}
\centerline{\scalebox{0.4}{\includegraphics{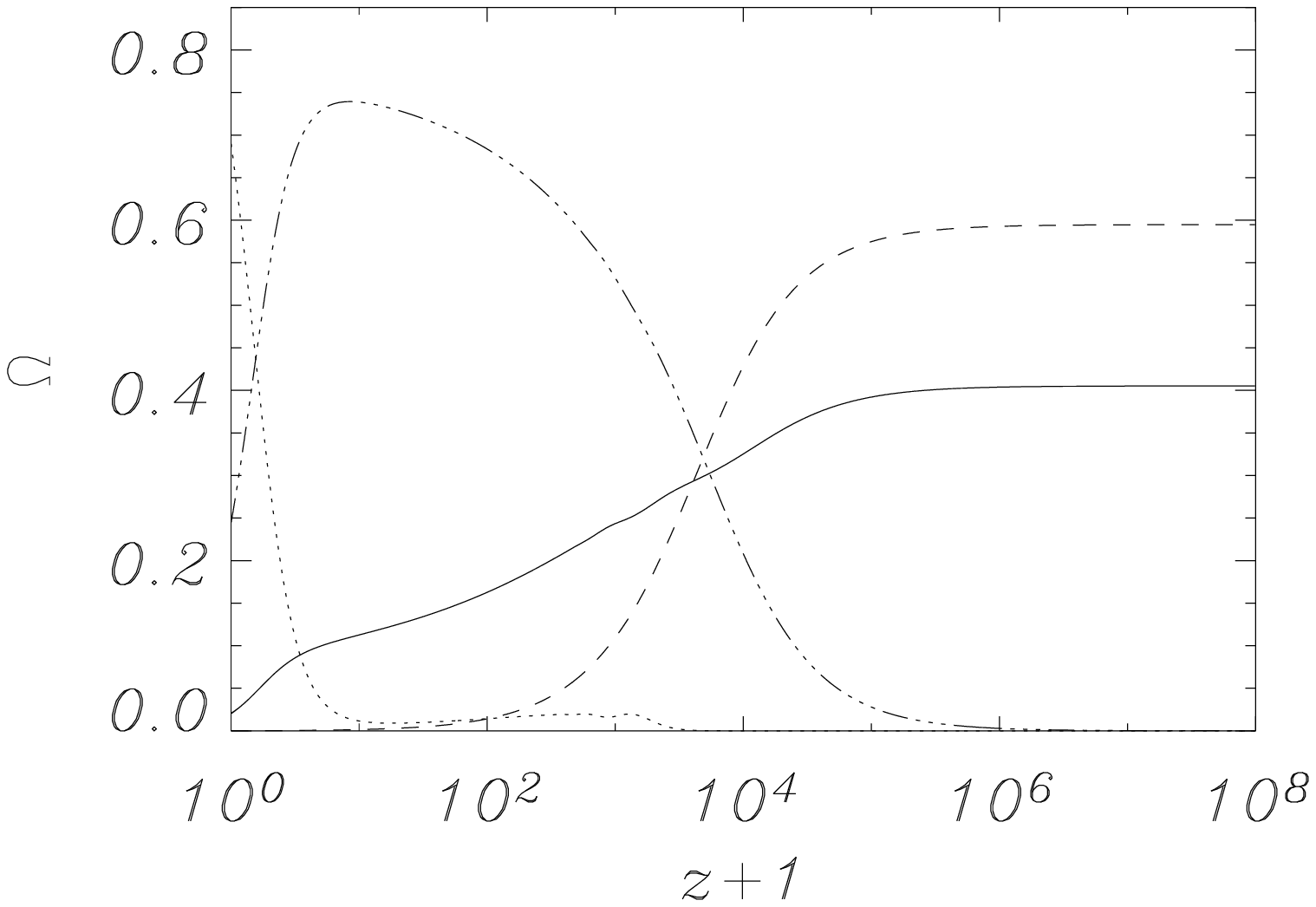}}}
\caption{Background evolution: In the upper panel, we plot the
evolution of the density parameters for a model with $\beta=0$,
$\lambda=1$. In the middle panel the corresponding plot with $\beta=1$
is shown, whilst the lower plot shows the case $\beta=-0.79$,
$\lambda=1$. (Neutrinos: solid line, CDM: dot-dashed line, scalar
field: dotted line and radiation: dashed line.) In all cases, the mass
of the neutrinos is $m_\nu = 0.314$~eV today. We consider a flat
universe with $\Omega_b h^2 = 0.022$, $\Omega_c h^2 = 0.12$,
$\Omega_\nu h^2 = 0.01$ and $h=0.7$.}
\end{figure}

For the model described in this section the coupled neutrinos are heavier in the past than
uncoupled neutrinos, which implies that the energy density stored in
the neutrinos is larger than would normally be expected. This means
that the evolution of the neutrino density parameter $\Omega_\nu$
depends on the evolution of the neutrino mass, which in turn depends
on the choice of the coupling $\beta$ and the slope of the potential
$\sigma$.  This can be seen from Figure 2. The coupling of neutrinos
to dark energy significantly alters the evolution of the cosmological
background. In particular it can be seen that the extra energy stored
in the neutrinos in the past can alter the redshift of
matter--radiation equality.

Typically one expects non--relativistic neutrinos to behave in a
similar manner to CDM, however the interaction between the neutrinos
and the scalar field modifies the scaling behaviour of the
non-relativistic neutrinos. It can be seen that the neutrino energy
density dilutes away faster than that of CDM, which is especially
notable for large values of coupling.  The evolution of $\phi$ caused
by the transfer of energy between the coupled neutrinos and scalar
field also results in the energy density of the quintessence field
becoming dominated by kinetic energy.  Finally at a redshift of the
order of unity the potential energy of the scalar field begins to
dominate, and all other matter species decay away.

A final point we would like to raise is the fact the {\it apparent}
equation of state measured by an observer is not given by
\begin{equation}
w_\phi = \frac{p_\phi}{\rho_\phi},
\end{equation}
where $\rho_\phi$ and $p_\phi$ are defined in eqns (\ref{rhop}) and
(\ref{pp}). One of the usual assumptions made in the measurements of
the dark energy equation of state using supernovae is that matter
(dark, baryonic or neutrinos) is decoupled from dark energy. At low
redshifts, all these components are assumed to scale like
$a^{-3}$. This is clearly not the case with the coupled neutrinos
here. It was pointed out in \cite{khoury}, that the apparent equation
of state is given by\footnote{The authors of \cite{khoury} called this
quantity an effective equation of state. However, we will define an
effective equation of state below.}
\begin{equation}\label{eq:wapp}
w_{\rm ap} = \frac{w_\phi}{1-x},
\end{equation}
with
\begin{equation}
x = -
\frac{\rho_{\nu,0}}{a^3\rho_\phi}\left[\frac{m_\nu(\phi)}{m_\nu(\phi_0)}
- 1\right].
\end{equation}
In this equation, the subscript $0$ denotes the quantities at the
present epoch. We emphasise that this quantity is {\it not} the
effective equation of state of dark energy, which is defined as
\begin{equation}
\dot \rho_\phi + 3H \rho_\phi\left(1+ w_{\rm eff}\right) = 0,
\end{equation}
while assuming that the neutrino density and neutrino pressure evolve
according to Eq. (\ref{eq:nuenergy}).  Using the Klein-Gordon
equation, one finds that the effective equation of state can be
written as
\begin{equation}
w_{\rm eff} = w_\phi +
\frac{\beta\dot\phi}{3H}\frac{\rho_\nu}{\rho_\phi}.
\end{equation}

In Fig. 3, we plot the apparent equation of state $w_{\rm ap}$ as a
function of redshift. Note that $w_{\rm ap}$ can be less than $-1$, as
was pointed out in \cite{wandelt}, \cite{amendola} and \cite{khoury}
in the context of models with dark matter/dark energy interaction.
\begin{figure}
\centerline{\scalebox{0.5}{\includegraphics{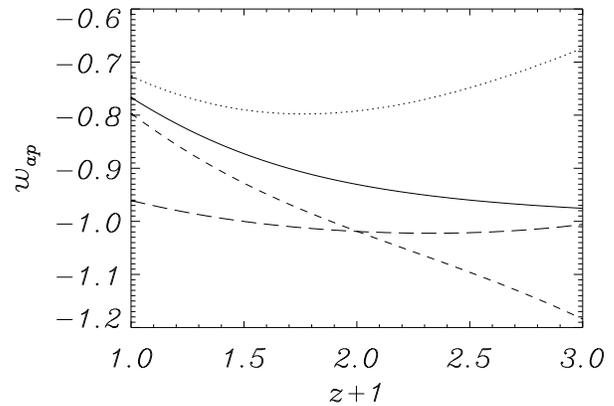}}}
\caption{ The apparent equation of state, as defined in
eq. (\ref{eq:wapp}), as a function of redshift $z$ (solid line:
$\beta=0$, $\lambda=1$; short dashed line: $\beta=1$, $\lambda = 1$;
dotted line: $\beta=-0.79$, $\lambda = 1$; long dashed line $\beta =
1$, $\lambda = 0.5$.)}
\end{figure}

To conclude this part, even if dark energy couples only to a
subdominant component such as neutrinos, the apparent equation of
state can be less than $-1$, without introducing phantom fields. As it
can be seen from Figure 3, the apparent equation of state might even
cross the boundary $w=-1$.

\section{Perturbation evolution}

Let us now turn our attention to the evolution of cosmological
perturbations in our model.  We adopt the conventions of Ma and
Bertschinger \cite{Ma} and work in the synchronous gauge, taking the
line element to be
\begin{equation}
ds^2=-a^2d\tau^2+a^2\left(\delta_{ij}+h_{ij}\right)dx^idx^j.
\end{equation}
(For a review of cosmological perturbation theory, see \cite{feldman},
\cite{durrer} or \cite{giovan}.)

As already mentioned in the last section, to fully describe the
evolution of cosmological neutrinos, we must calculate their
distribution function $f\left(x^i,p^i,\tau\right)$ in phase space.  We
can treat the neutrinos as collisionless particles throughout the
period of interest, and hence we can find the neutrino distribution
function by solving the collisionless Boltzmann equation \cite{Ma}

\begin{equation}\label{eq:cbe}
\frac{\partial f}{\partial\tau} + \frac{d x^i}{d\tau }\frac{\partial
f}{\partial x^i} + \frac{dq}{d\tau }\frac{\partial f}{\partial q} +
\frac{d n^i}{d\tau }\frac{\partial f}{\partial n^i} = 0,
\end{equation}
where the comoving momentum is $q_i = ap_i$.  It is convenient to
rewrite the comoving momentum in terms of its magnitude and
direction: $q^i = qn^i$.  The last term in equation (\ref{eq:cbe}) is
a second order quantity and will be neglected in the following linear
perturbation formalism.

The path of a neutrino in spacetime is governed by the general action
\begin{equation} \label{eq:action}
S = - \int m_\nu(\phi) \sqrt{-ds^2},
\end{equation}
which can be minimised to derive the neutrino geodesic equation\footnote{Please note Errata at the end of this paper.}
\begin{equation}
P^0\frac{\partial P^\rho}{\partial \tau}+\Gamma^\rho_{\alpha\beta}P^\alpha
P^\beta = -m_\nu^2 \frac{d\ln m_\nu}{d\phi}\frac{\partial \phi}{\partial
x_\rho},
\end{equation}
where $P^\mu$ is the proper momentum of the neutrino. Taking the
zeroth component of this equation and using the relation $P^0 =
\epsilon a^{-2}$, one finds that the comoving three-momentum of the
neutrinos is given by$^{3}$
\begin{equation}
\frac{dq}{d\tau}=-\frac{1}{2}q\dot{h}_{ij}n_i n_j.
\end{equation}
This equation does not depend explicitly on the coupling or the scalar
field perturbations.  Following \cite{Ma}, we write the phase space
distribution of the neutrinos as a zeroth order distribution plus a
small perturbation
\begin{equation}
f(x^i,p_j,\tau)=f_0(q)\left[1+\Psi\left(x^i,q,n_j,\tau\right)\right].
\end{equation}
Substituting this expression into the Boltzmann equation and
performing a Fourier transformation, we find$^{3}$
\begin{eqnarray}\label{eq:ftcbe}
\frac{\partial \Psi}{\partial \tau} &+& i\frac{q}{\epsilon}\left({\bf
k\cdot n}\right)\Psi \nonumber \\ &+& \frac{d\ln f_0}{d\ln q}
\left[\dot\eta-\frac{\dot h+6\dot \eta}{2}\left({\bf k\cdot
n}\right)^2\right]=0
\end{eqnarray}
In this equation (and in eqs. (\ref{pertkg}) and (\ref{denscont}), given
later), $\eta$ and $h$ are the standard scalar parts of the metric
perturbation $h_{ij}$. It is clear that equation
(\ref{eq:ftcbe}) does not contain terms proportional to $d\ln
m_\nu/d\phi$. Therefore, the equations for the neutrino hierarchy
derived in \cite{Ma} do not change$^{3}$. However, the expressions for the
perturbed neutrino energy density and neutrino pressure, which will be
calculated using $f$, are modified.  The perturbed energy density is
given by
\begin{equation}
\delta\rho_\nu = \frac{1}{a^4}\int q^2dq\,d\Omega f_0 \left(\epsilon\Psi + \frac{d\ln
  m_\nu}{d\phi}\frac{m_\nu^2a^2}{\epsilon}\delta\phi\right)
\end{equation}
which can be written as
\begin{equation}
\delta\rho_\nu=\frac{1}{a^4}\int q^2dq\,d\Omega\, \epsilon f_0(q)\Psi
+ \delta\phi \frac{d\ln m_\nu}{d\phi}(\rho_\nu-3p_\nu).
\end{equation}
Similarly, the expression for the perturbation in the neutrino
pressure is given by
\begin{equation}
\delta p_\nu = \frac{1}{3a^4}\int q^2 dq\,d\Omega
f_0(q)\left(\frac{q^2}{\epsilon}\Psi - \delta \phi \frac{d\ln m_\nu}{d
\phi} \frac{q^2 m_\nu^2 a^2}{\epsilon^3} \right)\nonumber.
\end{equation}
The expressions for the neutrino shear and energy flux remain
unchanged as they do not depend explicitly upon $m_\nu$.  Finally, the
perturbed Klein-Gordon equation is given by:
\begin{eqnarray}\label{pertkg}
\lefteqn{\ddot{\delta \phi}+ 2H \dot{\delta \phi}+\left(k^{2} +
a^{2}\frac{d^{2}V}{d\phi^{2}}\right)\delta \phi+\frac{1}{2}\dot{h}
\dot{\phi}=} \\ \nonumber & &-a^2 \left[\frac{d\ln
m_\nu}{d\phi}(\delta\rho_\nu-3\delta p_\nu)+\frac{d^{2} \ln
m_\nu}{d\phi^{2}}\delta\phi(\rho_\nu-3 p_\nu)\right].
\end{eqnarray} 

To calculate the temperature anisotropy spectrum and matter power
spectrum we apply these modifications to CAMB \cite{camb}. This code
calculates the linear cosmic background anisotropy spectra by solving
the Boltzmann equation which governs the evolution of the density
perturbations, and integrating the sources along the photon past light
cone.  To ensure the accuracy of our calculations, we directly
integrate the neutrino distribution function, rather than using the
standard velocity weighted series approximation scheme. We do not
consider lensing effects, nor tensor contributions.

\begin{figure}
\centerline{\scalebox{0.5}{\includegraphics{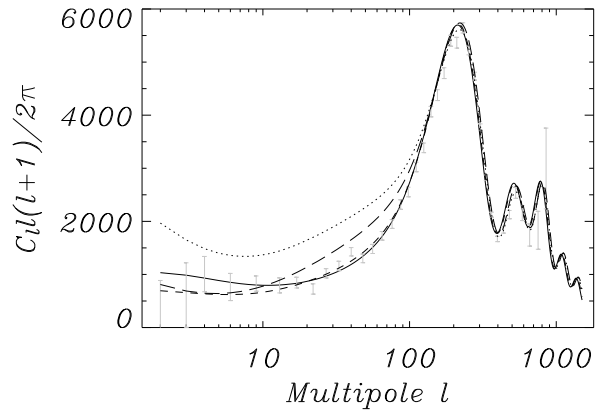}}}
\caption{The CMB anisotropy spectrum (unnormalized) for exponential
coupling and potential. Solid line: $\beta=0$, $\lambda = 1$;
short--dashed line: $\beta = 1$, $\lambda=1$; dotted line:
$\beta=-0.79$, $\lambda = 1$; long--dashed line: $\beta=1$,
$\lambda=0.5$. Error bars denote WMAP data.}
\end{figure}

\begin{figure}
\centerline{\scalebox{0.5}{\includegraphics{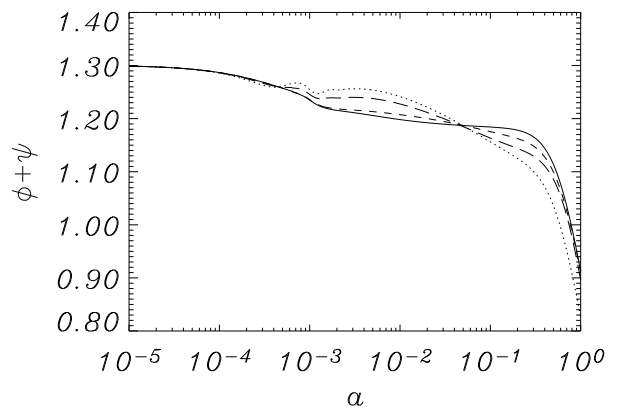}}}
\caption{Evolution of the sum of the metric perturbations $\Phi+\Psi$.
Solid line: $\beta=0$, $\lambda = 1$; short--dashed line: $\beta = 1$,
$\lambda=1$; dotted line: $\beta=-0.70$, $\lambda = 1$; long--dashed
line: $\beta=1$, $\lambda=0.5$.  The scale is $k=10^{-3}$Mpc$^{-1}$.}
\end{figure}

\begin{figure}
\centerline{\scalebox{0.5}{\includegraphics{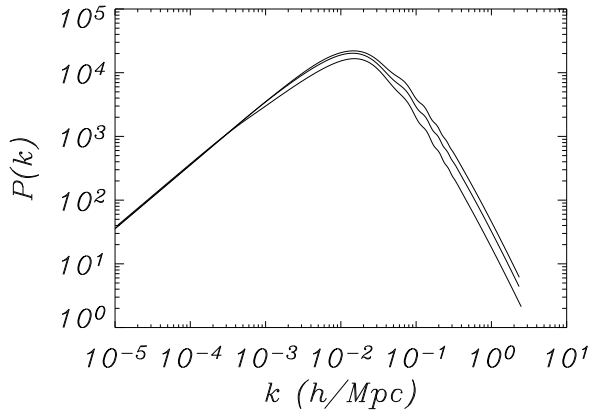}}}
\caption{Plot of the matter power spectrum. From the top curve to the
bottom curve: ($\beta=0$, $\lambda=1$), ($\beta=1$, $\lambda=0.5$),
($\beta=-0.79$, $\lambda=1$). The matter power spectrum for
($\beta=1$, $\lambda=1$) is indistinguishable from the ($\beta=0$,
$\lambda=1$) curve.}
\end{figure}

The results of the neutrino-dark energy coupling on the temperature
anisotropy spectrum can be seen in Figure 4.  The most obvious
modifications to the anisotropy spectrum occur for large angular
scales, with $\ell<100$, although for some choices of parameters the
positions and relative heights of the peaks are also affected.  We
generally observe an increase in power for $10<\ell<100$, whilst for
$\ell<10$ we find either an excess or reduction in power depending
upon our choice of parameters. Note that this is in marked contrast to
models of coupled CDM, where an increase in power on large scales is
usually observed \cite{domenico}. For the models where the neutrinos
were much heavier in the past than today, we also observe a slight
shift in the acoustic peaks and a change in their relative amplitudes.

For scales larger than a degree ($\ell<100$), the dominant
contribution to the anisotropy spectrum is the Integrated Sachs-Wolfe
Effect (IWS).  This arises due to the evolution of the gravitational
potentials along the photon path from the surface of last scattering.
The modification to the cosmological background arising from the
neutrino coupling can have a significant effect upon the evolution of
the perturbations.  In particular there is a larger energy density in
neutrinos in coupled models during the transition period when the
neutrinos become non--relativistic.  As a result, the intermediate
regime between radiation and matter domination is prolonged, and so
the evolution of the gravitational potentials are significantly
modified. The evolution of the sum of the metric perturbations $\Phi$
and $\Psi$ is shown in Figure 5. The modifications to the behaviour of
the metric perturbations for the different models is immediately
apparent.

For very large scales ($\ell\leq20$) anisotropies arise primarily from
the late time Integrated Sachs-Wolfe effect (ISW), which is caused by the evolution of the metric
perturbations for redshifts in the range $0<z<2$.
In particular, $\rho_\phi$ and $\rho_\nu$ as well as the equation of
state of dark energy affect the late time behaviour of cosmological
perturbations.  As mentioned above, the evolution of the scalar field
is influenced by the presence of the coupling to the neutrinos and
hence the equation of state of dark energy depends upon
$\beta$. Likewise, the clustering properties of dark energy depends on
the neutrino coupling (see \cite{weller} for a discussion on the
clustering of dark energy and its impact on the CMB).  The neutrinos
will generally tend to fall into the potential wells of dark matter,
although at a rate slightly dependent on the coupling to the scalar
field. The scalar field itself will cluster together with the
neutrinos and thereby affecting the gravitational potential.

Let us turn our discussion to the evolution of neutrino
perturbations. Figure 6 shows the effects of neutrino coupling on the
matter power spectrum. Here we typically observe damping, and our
results appear similar to standard models of CDM and hot dark matter,
where a similar reduction in power could be achieved with a heavier
neutrino mass.

We can use the perturbed part of the energy momentum conservation
equation for the coupled neutrinos
\begin{equation}
T^\mu_{~~\gamma;\mu} =\frac{d \ln m_\nu}{d\phi}
\phi_{,\gamma}T^{\alpha}_{~\alpha}
\end{equation}
to calculate the evolution equations for the neutrino perturbations
($T^{\alpha}_{~\alpha}$ stands for the trace of the neutrino energy
momentum tensor and the semicolon denotes the covariant derivative).
Taking $\gamma=0$ we derive the equation governing the evolution of
the neutrino density contrast , $\delta_\nu \equiv \frac{\delta
\rho_\nu}{\rho_\nu}$ whilst taking $\gamma=i$ (spatial index) yields
the velocity perturbation equation $\theta_\nu \equiv
ik_{i}v^{i}_{\nu}$, with the coordinate velocity $v^{i}_{\nu} \equiv
dx^{i}/d {\tau}$:
\begin{eqnarray}\label{denscont}
\dot{\delta}_\nu &=& 3\left(H+\beta
\dot{\phi}\right)\left(w_\nu-\frac{\delta p_\nu}{\delta
  \rho_\nu}\right)\delta_\nu - \left(1+w_\nu\right)\left(\theta_\nu +
\frac{\dot{h}}{2}\right)\nonumber \\ &+&
\beta\left(1-3w_\nu\right)\dot{\delta
  \phi}+\frac{d\beta}{d\phi}\dot{\phi}\delta\phi\left(1-3w_\nu\right),
\end{eqnarray}

\begin{eqnarray}
\dot{\theta_\nu} &=& -H(1-3w_\nu)\theta_\nu -
\frac{\dot{w_\nu}}{1+w_\nu}\theta_\nu+\frac{\delta p_\nu / \delta
\rho_\nu}{1+w_\nu}k^{2}\delta_\nu \nonumber \\ &+&
\beta\frac{1-3w_\nu}{1+w_\nu} k^{2} \delta
\phi-\beta(1-3w_\nu)\dot{\phi}\theta_\nu - k^{2} \sigma_\nu.
\end{eqnarray}
The variable $\sigma_\nu$ represents the neutrino anisotropic stress
and we have used the more general definition $\beta = d\ln
m_\nu/d\phi$, which in general might be not constant. Furthermore the
neutrino equation of state is given by $w_\nu \equiv p_\nu /
\rho_\nu$.

It is the presence of the additional coupling terms in these
expressions for the growth of the neutrino density and velocity
perturbations, as well as the modifications to the evolution of the
cosmological background, which alters the behaviour of the neutrino
perturbations in comparison with the standard uncoupled case.

\begin{figure}\label{fig:denconk0p1}
\centerline{\scalebox{0.5}{\includegraphics{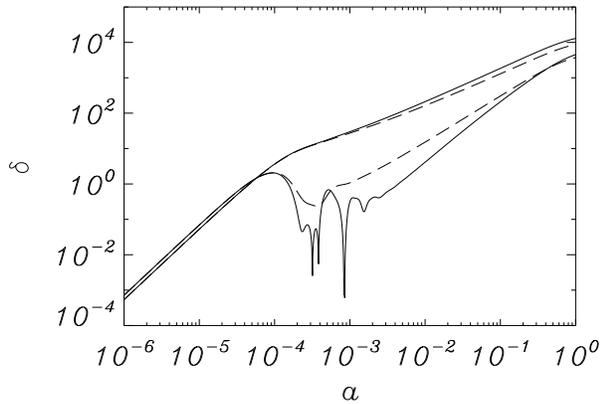}}}
\caption{Evolution of the neutrino (lower curves) and CDM (upper
curves) density contrasts. The solid line shows the uncoupled case,
i.e. $\beta=0$, whereas the dashed line shows the case for
$\beta=-0.79$.  The scale is $k=0.1$Mpc$^{-1}$.}
\end{figure}

Figure 7 shows the evolution of the neutrino and CDM density
contrasts, comparing the uncoupled model with an extreme coupled case
with $\beta=-0.79$, for which the mass of the neutrinos is $m_\nu
\approx 2.5$~eV at $z\ge 1400$ but $m_\nu=0.3$~eV today. Deep inside
the radiation dominated epoch the neutrinos are highly relativistic
and their density contrast grows in a similar manner to radiation.  In
the uncoupled models the growth of the density perturbations of the
neutrinos makes a transition to matter--like behaviour once the
neutrinos become non--relativistic. The neutrinos will fall into the
CDM potential wells, which is the dominant component at
recombination. At small wavelengths (large wavenumbers), the neutrinos
undergo freestreaming, which prevents neutrinos from clustering at an
arbitrary small scale.  The freestreaming length scale after the
neutrinos become non-relativistic can be estimated to be \cite{Ma} 
(reinstating Newton's constant $G$)
\begin{equation}
k_{\rm fs}(a) = \frac{4\pi G \rho a^2}{v_{\rm med}^{2}}
\end{equation}
where $\rho$ is the background total density. The median neutrino
speed is given by:
\begin{equation}
v_{\rm med}=15 a^{-1} \left( \frac{m_\nu(a)}{10 \rm{~eV}} \right)
\rm{km~s^{-1}}.
\end{equation}
Since the neutrino momentum decays like $a^{-1}$, the neutrino
velocity behaves like $\left(a m_\nu(a) \right)^{-1}$, taking into
account that the neutrino mass evolves with time.  Freestreaming stops
as soon as $k<k_{\rm fs}$, allowing the neutrino density contrast to
grow. This behaviour can clearly be seen in Figure 7 for both the
uncoupled and coupled cases.  At around $z\approx 10^4$ the neutrinos
become non--relativistic and start to freestream immediately, as can
be seen from the oscillating behaviour of $\delta_\nu$. At this stage
$k_{\rm fs}<k$. However, as soon as $k_{\rm fs}=k$ freestreaming
stops, and $\delta_\nu$ can grow unimpeded. 
The case with neutrino-dark energy coupling differs from the uncoupled
case since in the result shown the neutrinos are heavier in the past,
so for a given redshift $k_{\rm fs}$ is larger. This means that
freestreaming stops earlier than in the uncoupled case. This behaviour
is apparent in Figure 7 (dashed lines), where we see that $\delta_\nu$
starts to grow earlier than in the uncoupled case (solid lines).  The
neutrino-coupling also has an effect on the growth of the density
contrast itself since we observe that $\delta_\nu$ grows slower than
in the uncoupled case.  This is probably because the rate of
gravitational infall of the neutrinos tends to be reduced by the
presence of the much less clustered dark energy. Also, the coupling
has a slight effect on the growth of the dark matter density contrast,
which arises from the fact that the background evolution is modified.

\section{Another choice of coupling and potential}

So far we have restricted our discussion to one choice of
quintessence--neutrino coupling and one form for the dark energy
potential.  At this stage, the reader might wonder whether the results
obtained so far are simply due to our choice of potential and
coupling.  For a scalar field with standard kinetic term, the
exponential potential is not a favored model for a quintessential
potential, since the initial value of the scalar field has to be
fine--tuned to obtain scalar field domination today
\cite{copelandwands}. The interesting alternative possibility of a
global attractor unfortunately does not seem to be viable due to the
large perturbation growth \cite{domenico}. The coupling of the scalar
field to neutrinos does not cure the fine-tuning problem of the
exponential potential.

For our second form of coupling we choose
\begin{equation}
m_\nu (\phi) = M_0 e^{\beta\phi^2},
\end{equation}
which was also recently used in a model with dark matter/dark energy
coupling in \cite{massimo}.  The effect of this choice is that the
coupling function $d\ln m_\nu/d\phi$ becomes {\it field--dependent},
whereas it has been constant so far. Depending on how the field
evolves with time, the coupling can either grow or become smaller
during the cosmic history. Field--dependent couplings are not uncommon
in higher--dimensional theories and can appear in brane--world
theories, see for example \cite{brax1} or \cite{brax2}.

For the potential, we choose an inverse power-law potential, which is
a well--motivated candidate for a quintessence field (see
e.g. \cite{steinhardt} and \cite{binetruy}). To be concrete, we use
\begin{equation}
V(\phi) = \frac{M^6}{\phi^2}.
\end{equation}
With these choices for the potential and coupling, the effective
minimum will exist if $\beta>0$.

The results for the neutrino--mass evolution, the apparent equation of
state and the CMB anisotropy power-spectrum are shown in Figures 8, 9
and 10.
\begin{figure}
\centerline{\scalebox{0.5}{\includegraphics{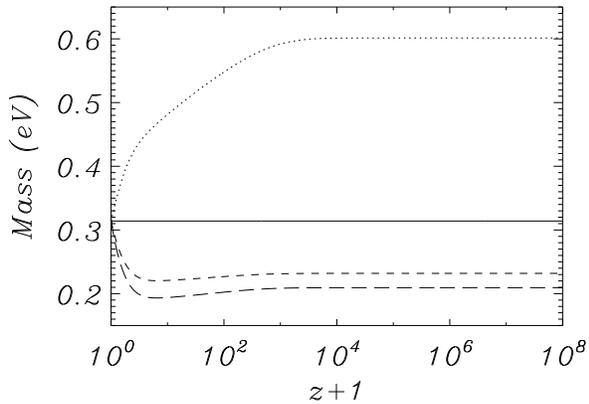}}}
\caption{The evolution of the neutrino mass for the inverse power-law
potential with $m_\nu(\phi)=m_0e^{\beta \phi^2}$ (solid line:
$\beta=0$; short dashed line: $\beta=0.2$; dotted line: $\beta=-0.2$;
long dashed line $\beta = 0.27$.)}
\end{figure}
\begin{figure}
\centerline{\scalebox{0.5}{\includegraphics{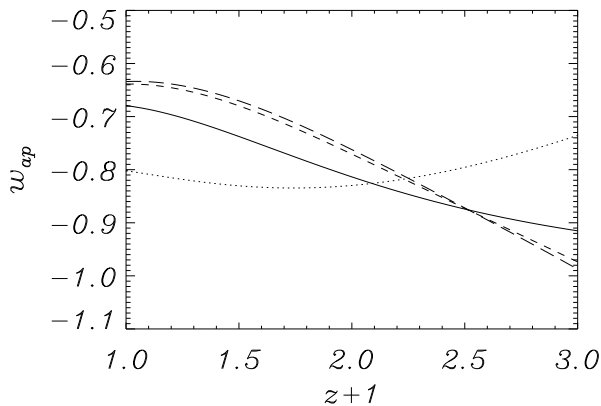}}}
\caption{The evolution of the apparent dark energy equation of state
for the inverse power-law potential with $m_\nu(\phi)=m_0e^{\beta
\phi^2}$ (solid line: $\beta=0$; short dashed line: $\beta=0.2$;
dotted line: $\beta=-0.2$; long dashed line $\beta = 0.27$.)}
\end{figure}

\begin{figure}
\centerline{\scalebox{0.5}{\includegraphics{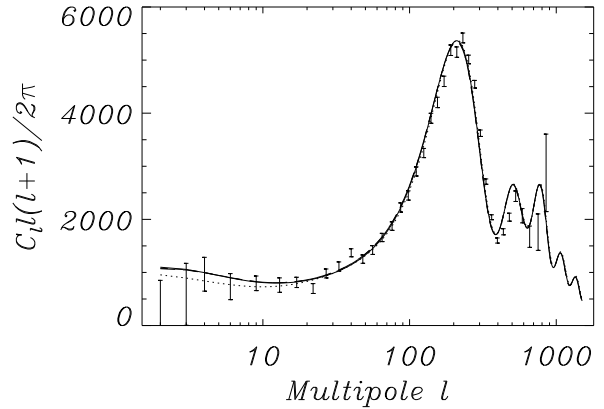}}}

\caption{The CMB anisotropy spectrum (unnormalized) for an exponential
potential and $m_\nu(\phi)=m_0e^{\beta \phi^2}$ (solid line:
$\beta=0$; short dashed line: $\beta=0.2$; dotted line: $\beta=-0.2$;
long dashed line $\beta = 0.27$.)}
\end{figure}

The biggest difference to the case of a purely exponential coupling is
that for positive $\beta$ the neutrinos are lighter in the past, as
can be seen from Figure 8. Thus, with a convenient choice of potential
and coupling, the neutrinos can become heavier as the universe
expands. In the case of a negative $\beta$, the effective potential
does not have a minimum and the neutrinos become lighter as the
universe expands. The results for the apparent equation of state are
shown in Figure 9. The results are similar to the ones found in
Section II: $w_{\rm ap}$ can be smaller than $-1$ and can cross the
boundary of the cosmological constant with $w=-1$.  As it can be seen
in Figure 9, the apparent equation of state varies substantially in
the redshift range $z=0-2$ if $\beta$ is non-zero. A strongly, however,
varying equation of state is not preferred by the data.  Finally, the
effects on the CMB anisotropies are similar to the ones found in
Section III, as can be seen from Figure 10. The only visible deviation
from $\beta=0$ is the case with negative $\beta$, in which a reduction
of power at low multipoles can be observed. The cases with positive
$\beta$ can not be distinguished from the uncoupled case. The reason
is that the neutrino density is smaller in the past than in the
uncoupled case for this choice of potential and coupling. Hence,
neutrinos are less important for the dynamics of the universe, and
their imprint upon the CMB is correspondingly reduced.

In essence, the physical explanations of the model presented in our
earlier paper \cite{us} and in Sections II and III remain valid even for
other choices of the potential and couplings, since they show how to
relate the general behaviour of a dynamical neutrino mass to the
cosmological evolution.

\section{Parameter Constraints}

In the earlier sections we demonstrated that models of coupled dark
energy and neutrinos could produce a detectable signature in
cosmological surveys.  Indeed, the modifications to the background
evolution (mainly to the dark energy equation of state), temperature
anisotropy spectrum and matter power spectrum should allow us to
constrain our model using current data sets.

We perform our likelihood analysis using CosmoMC \cite{cosmomc}.  This
program uses a Markov--Chain Monte--Carlo (MCMC) engine to efficiently
explore the cosmological parameter space.  Typically we run five
chains for each simulation, with no less than 35,000 samples per
chain. We perform the usual convergence checks on the individual
chains to ensure that the chains have fully sampled our parameter
space.  As well as visually confirming that the individual chains
converge, we check the Gelman and Rubin R statistic (variance of chain
means/mean of chain variances) for each parameter, and ensure that the
Raftery and Lewis convergence diagnostic is satisfied.

We take advantage of the wide range of cosmological data which is
currently available to constrain our model.  The CMB temperature
anisotropy spectrum is constrained using WMAP \cite{wmap1}
\cite{wmap2}, CBI \cite{cbi}, ACBAR \cite{acbar} and VSA \cite{vsa}
datasets. The neutrino--dark energy coupling can also affect the
formation of large scale structure which is sensitive to the neutrino
mass, and so we use data from the Sloan Digital Sky Survey \cite{sdss}
to further constrain our model. Data from the Supernova Cosmology
Project \cite{supernova} can also be used to constrain the equation of
state of dark energy, and thus place further constraints on our model.

We choose to perform the data analysis using our usual choice of
potential and coupling, namely $V(\phi)=V_0e^{-\sigma\phi}$ and
$m_\nu(\phi)=M_0e^{\beta\phi}$. We choose to focus on these potentials
because they embody the typical behaviour observed for most models of
coupled neutrinos, and they easily reduce to the standard $\Lambda$CDM
case ($\beta=\sigma=0$).  These potentials also have the advantage
that the initial choice of $\phi_i$ does not affect the evolution of
the cosmology, as changes to the initial choice of $\phi$ are
equivalent to re-scalings of the mass parameters $M_0$ and $V_0$.  For
general choices of potentials and couplings this useful degeneracy
does not exist, as the neutrino--dark energy coupling can severely
restrict the range of attractor solutions. Consequently the increased
number of fine-tuned free parameters required for these models would
compromise the goodness of fit compared to simpler models.

\begin{figure}
\centerline{\scalebox{0.27}{\includegraphics{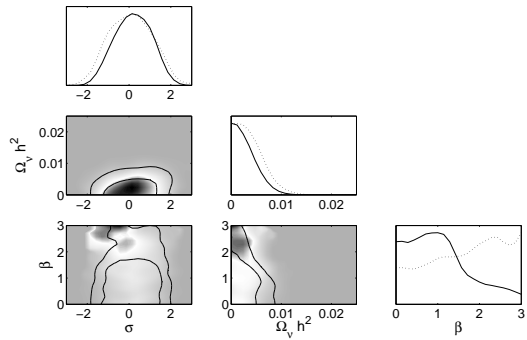}}}
\caption{Posterior constraints for 9 parameter model described in the
text.  Shading denotes the mean likelihood of the samples, whilst the
contours show the 68\% and 95\% confidence limits from the
marginalised distribution.  Solid lines on the 1-D plots show the
marginalised posterior, whilst the dotted curves denote the mean
likelihood of the samples.}
\label{betasigmafnu}
\end{figure}

Throughout our analysis we assume a flat universe, with $\Omega=1$.
Initially we use the standard parameterization for our cosmological
model, and vary the following parameters: $\Omega_{b} h^2$,
$\Omega_{\rm CDM}h^2$, $h$, $z_{\rm re}$ (the redshift of
reionisation), $\Omega_\nu h^2$, $n_s$ (the spectral index), $10^{10}A_s$ 
(the initial scalar perturbation amplitude) and the dark energy parameters 
$\sigma$ and $\beta$.  We show the results from this initial analysis is Figure
\ref{betasigmafnu}.

Clearly Figure \ref{betasigmafnu} shows that current cosmological data
places no constraints on our new coupling parameters.  It is well
known that the current best fit analysis of cosmological data can only
place an upper limit on the mass of the neutrino, and that massless
neutrinos are not excluded by most cosmological data sets.  However
the results presented so far in this paper require that the neutrinos
having a significant mass; indeed for low mass neutrinos the effects
observed in this paper become largely insignificant.
Despite this, as we will show later, the strength of the neutrino
coupling can still be constrained by the requirement for later time
acceleration.

Current cosmological data requires that the universe contains
approximately 70\% of dark energy, 30\% of dark matter and some minor
quantities of baryons and neutrinos. In the dynamics of our model
there could exist \emph{only} one critical point able to guarantee
such proportions in its vicinity (see e.g. \cite{domenico}).  When
exactly reached, this point is characterized by the total domination
of the scalar field and exists only for $\lvert \sigma \rvert < \sqrt{6} \approx
2.5$. Indeed our computed likelihood will be shown to be contained in
such boundaries. A preferred range for the parameter $\beta$ is more
difficult to predict, however, because of the previously described
effects induced on the background and perturbation at different stages
of the evolution, although it is clear that very large values of
$\beta$ will not be favoured by the data. But it is very important to
emphasize that values of $\beta$ of the order of unity are perfectly
acceptable.

Another consideration when choosing the scalar field parameters is the
BBN constraint resulting from the early time modification of the scale
factor 
evolution due to the presence of dark energy \cite{copeland,bean}. 
As discussed earlier, the neutrino coupling to dark energy in our model 
does not modify the energy density of the dark energy at the time of 
BBN (when the neutrinos are highly relativistic, and the coupling terms 
are negligible). For a quintessence potential with a 
scalar field dominated
late time attractor 
the BBN constraints of \cite{bean} are easily satisfied. We have confirmed 
this numerically for our coupled models (for instance, in the case of $\beta=1$, $\Omega_{\phi}\sim 10^{-25}$ at $z\sim 10^8$). Note that, in principle, our 
parameter space includes both the late-time attractor and scaling solutions 
(that evolve like radiation or matter and 
which would provide a non-negligible contribution to the energy density at 
the time of BBN). We find, however, that the scaling solution for
the exponential potential is already strongly disfavored by the observations 
that we have used for our analysis and therefore additional BBN constraints 
would not modify our findings. 

Recent works \cite{bridlexray} have used data from X-ray clusters to
reduce the uncertainties on the lower bound for the neutrino mass,
finding a value for the neutrino masses of $\sum
m_\nu=0.56^{+0.30}_{-0.26}$~eV. Notice though that the upper bounds could 
in fact go up to $\sum m_{\nu}=2$~eV, depending 
on the datasets used and the assumed priors \cite{elgaroy}. Measurements of atmospheric 
neutrino oscillations suggest that there is at least one neutrino species with
$m_\nu>0.05$~eV \cite{massiveneutrinos1}. The Heidelberg-Moscow experiment, searching for neutrino-less double beta decay claim detection of
electron neutrino mass $m_{\nu_e}$  between $0.2$~eV and $0.6$~eV, with best fit $m_{\nu_e}=0.36$~eV \cite{moscow}.

We therefore choose to
perform our analysis using two values of the neutrino mass today,
$m_\nu=0.2$~eV and $m_\nu=0.3$~eV, to investigate whether models of
neutrino--dark energy coupling could in principle be constrained if
neutrinos were independently confirmed to have a significant ($m_\nu
\gtrsim 0.1$~eV) mass, consistent with current experiments measuring
the neutrino mass. 

By choosing to fix the value of the neutrino mass today, we are
required to specify the current value for the energy density stored in
neutrinos and value of the Hubble constant as the neutrino mass, the
critical energy density in neutrinos and the Hubble constant are
related via the usual formula
 \begin{equation}
 \Omega_\nu h^2 = \frac{\sum m_\nu}{93.2\text{~eV}}.
 \end{equation}

 We also choose to fix the value of $\Omega_b h^2$ as we do not expect
 the behaviour and constraints of the baryon energy density to be
 significantly modified by our neutrino--dark energy coupling, as the
 observed effects on the anisotropy spectrum are largely limited to
 relatively low multipoles.
 
 The values for $H_0$ and $\Omega_b h^2$ can be determined
 independently from the cosmological data used in our MCMC analysis.
 The value of $H_0=72~\rm{km~s^{-1}~Mpc^{-1}}$ can be obtained from the best
 fit of the Hubble Space Telescope Key Project \cite{hst}, whilst the
 baryon density parameter $\Omega_b h^2= 0.022$ is favoured by Big
 Bang Nucleosynthesis models \cite{bbn}. We are therefore left with a
 cosmological model requiring 6 parameters: $\beta$, $\sigma$,
 $\Omega_\text{CDM}h^2$, $z_{\rm re}$, $A_s$ and $n_s$.

\begin{figure}
\centerline{\scalebox{0.27}{\includegraphics{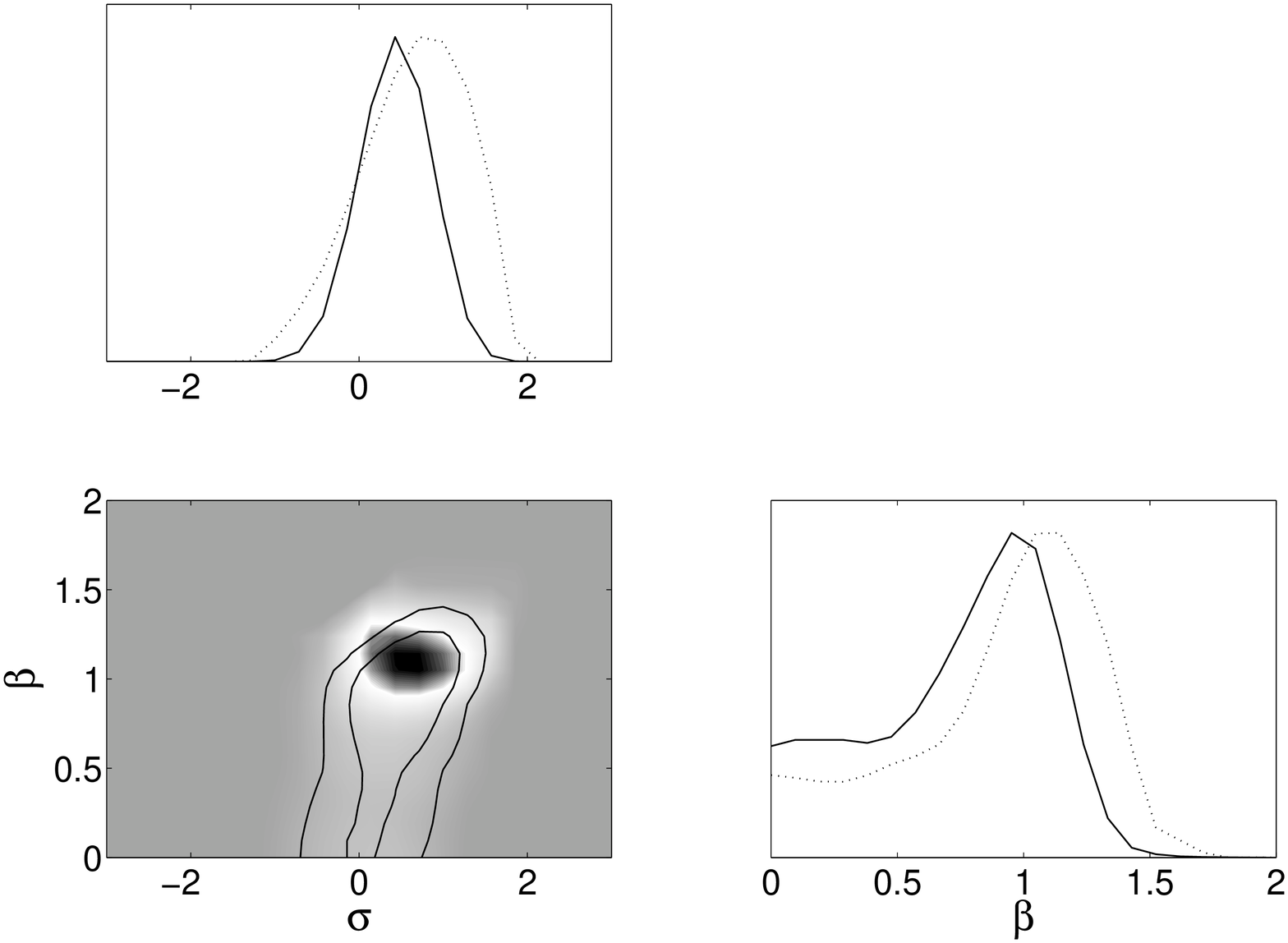}}}
\caption{Posterior constraints for 6 parameter model, with
$m_\nu=0.2$~eV, $\Omega_b h^2=0.022$ and $h=0.72$.  Shading denotes
the mean likelihood of the samples, whilst the contours show the 68\%
and 95\% confidence limits from the marginalised distribution.  Solid
lines on the 1-D plots show the marginalised posterior, whilst the
dotted curves denote the mean likelihood of the samples.}
\label{fig:betasigmaslow}
\end{figure}

\begin{figure}
\centerline{\scalebox{0.27}{\includegraphics{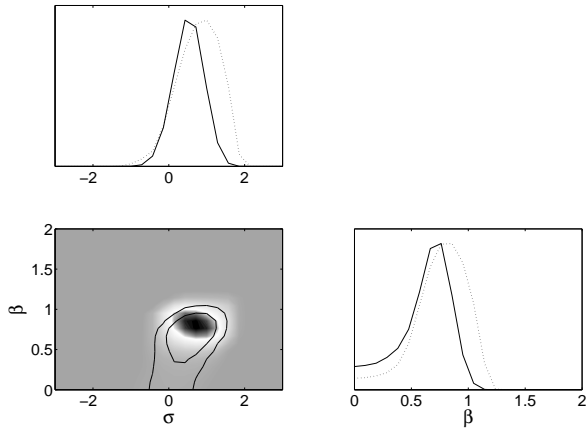}}}
\caption{Posterior constraints for 6 parameter model, with
$m_\nu=0.3$~eV, $\Omega_b h^2=0.022$ and $h=0.72$.  Shading denotes
the mean likelihood of the samples, whilst the contours show the 68\%
and 95\% confidence limits from the marginalised distribution.  Solid
lines on the 1-D plots show the marginalised posterior, whilst the
dotted curves denote the mean likelihood of the samples.}
\label{fig:betasigma}
\end{figure}

The results of the MCMC analysis for neutrinos with a mass today of
$m_\nu=0.2$~eV can be found in Table \ref{tab:lightnu}, whilst the
results for $m_\nu=0.3$~eV are given in Table \ref{tab:heavynu}.  We
quote the marginalised probability distributions and confidence
intervals.  Figures \ref{fig:betasigmaslow} \& \ref{fig:betasigma}
show the 2D probability distributions for the $m_\nu=0.2$~eV and
$m_\nu=0.3$~eV models respectively. 

%
\begin{table}
\begin{center}
\begin{tabular}{|c|c|c|c|}
\hline Parameter & Mean & 68\% & 95\% \\ & likelihood & interval &
interval \\ \hline \hline $\Omega_\text{CDM} h^2$ & $0.102\pm 0.004$ &
$0.099-0.106$ & $0.096-0.110$ \\ $z_\text{re}$ & $17.6\pm 3.7$ &
$16.0-19.8$ & $10.7-23.2$ \\ $\sigma$ & $0.43\pm 0.32$ & $0.29-0.60$ &
$0.13-0.95$ \\ $\beta$ & $0.75\pm 0.64$ & $0.64-0.98$ & $0.11-1.18$ \\
$n_s$ & $0.96\pm 0.01$ & $0.95-0.97$ & $0.93-0.99$ \\
\hline
\end{tabular}
\end{center}
\caption{Marginalised parameter constraints for our 6 parameter model
with fixed $m_\nu=0.2$~eV, $\Omega_b h^2=0.022$ and $h=0.72$. For this model we 
find $\chi^2/\text{dof} = 1570.1/1459$. This compares with a $\chi^2/\text{dof} = 1610.1/1461$ for a best-fit $\Lambda$CDM model using the same parameter set with $\sigma=\beta=0$.}
\label{tab:lightnu}
\end{table}
\begin{table}
\begin{center}
\begin{tabular}{|c|c|c|c|}
\hline Parameter & Mean & 68\% & 95\% \\ & likelihood & interval &
interval \\ \hline \hline $\Omega_\text{CDM} h^2$ & $0.100 \pm 0.003$
& $0.097-0.104$ & $0.094-0.107$ \\ $z_\text{re}$ & $20.5\pm 3.1$ &
$19.4-22.0$ & $15.0-25.1$ \\ $\sigma$ & $0.52\pm 0.29$ & $0.40-0.67$ &
$0.00-0.97$ \\ $\beta$ & $0.62\pm 0.21$ & $0.58-0.74$ & $0.15-0.86$ \\
$n_s$ & $0.97\pm 0.01$ & $0.96-0.99$ & $0.95-1.00$ \\
\hline
\end{tabular}
\end{center}
\caption{Marginalised parameter constraints for our 6 parameter model
with fixed $m_\nu=0.3$~eV, $\Omega_b h^2=0.022$ and $h=0.72$. In this case we find 
$\chi^2/\text{dof} = 1593.7/1459$. This compares with a $\chi^2/\text{dof} = 1636.8/1461$ for a best-fit $\Lambda$CDM model using the same parameter set with $\sigma=\beta=0$.}
\label{tab:heavynu}
\end{table}

As expected, the neutrino coupling has little effect on the value of
$\Omega_{CDM} h^2$ as the peak structure of the temperature anisotropy
spectrum is largely unaffected.  The value found for $n_s$ is also
within the usual range for parameter analysis of cosmological models
which neglect tensor contributions.

For both the $m_\nu=0.2$~eV and $m_\nu=0.3$~eV models we find that
non-zero values of neutrino coupling strengths are preferred by the
data.  We also see that for these models a non--zero value for
$\sigma$ is preferred over the usual cosmological constant, although
$\sigma=0$ is not excluded at the 68\% level.  This is not surprising
since we have seen that the equation of state of the dark energy for
some choices of parameters in our coupled models is entirely
consistent with the preferred value of $w_{ap}\sim -1$ found from
supernova surveys.  It is clear that models with heavier neutrinos
allow stronger constraints to be placed upon the strength of the
coupling.  Indeed, for the 0.3~eV neutrinos we find that
neutrino--dark energy coupling is preferred at the 1 sigma confidence
level.  This is to be expected as a larger neutrino mass today is
equivalent to a higher energy density in neutrinos, and so any
modification to the neutrino evolution will have a larger impact on
the CMB and LSS for models with greater densities of neutrinos.  In
particular we have seen that there exist a range of non-zero $\beta$
values capable of reducing power at low CMB multipoles.  This last
effect is most probably the cause for the relative peak in the
likelihood for $\beta$ of order unity. Furthermore a sharper drop at
large values of $\beta$ is observed in the likelihood most likely to
limit the excessive growth at multipoles $10 < \ell <100$.

It is however important to make clear that these constraints rely upon
the assumption that the neutrino mass is known, and that the neutrinos
have a mass $m_\nu \gtrsim 0.1$~eV. Although this assumption is
consistent with current neutrino experiments, we can only make the
statement that should the mass of the neutrino be found to be greater
than $0.1$~eV, then cosmological data can be used to constrain the
strength of any neutrino--dark energy coupling; indeed we find that
there is some evidence that the existence of such a coupling is
actually preferred by current cosmological data over the standard
$\Lambda$CDM cosmology.

\section{Conclusions}

We have investigated models of dark energy which couple a quintessence
scalar field to massive neutrinos.  In these models, dark energy and
neutrinos are coupled such that the neutrino masses become functions
of the scalar field.  The effects of such models on the cosmological
background evolution, on the cosmic microwave background anisotropies,
and on the formation of large scale structures were
analyzed. Additionally, we have also performed a likelihood analysis
on the parameter space of such theories.

We have focused on two specific models: In the first, the coupling
between neutrinos and dark energy is constant and the quintessential
potential is an exponential. The second model, which is better
motivated from the particle physics point of view, has a
neutrino--coupling which depends on the quintessence field (hence
changes with time), whilst the scalar field has a power-law potential.
In spite of some specific differences between these two models (such as 
the energy density stored in the scalar field at recombination for example), 
the effects of the coupling on the CMB anisotropies and on the matter 
power spectrum are nevertheless explainable by the basic mechanisms that we have identified
earlier. Namely, the coupling modifies the background history and
induces an ISW contribution to the CMB spectrum; the matter power
spectrum is modified by the magnitude of the neutrino mass during
structure formation. Given the generality of these explanations, the
conclusions drawn from this investigation could probably be applied to
any similar model with a neutrino-dark energy coupling.

It is important to note that in our models, the dark energy sector is
described by a \emph{light} scalar field, with a mass which is at most
of order $H$.  This is in contrast to previous models \cite{fardon1}
in which the mass of the scalar field is much larger than $H$ for most
of its history. The latter can have significant effects upon the
behaviour of the neutrinos and the growth of their perturbations, and
which is difficult to reconcile with current astronomical data
\cite{zalda}.

Solving the collisionless Boltzmann equation for the neutrinos, we
have investigated the relativistic and non-relativistic regimes and
the transition period in between.  Initially the neutrinos are highly
relativistic, and during this period the quintessence field is
frozen. The mass of the neutrinos therefore remains constant. As the
neutrinos become non-relativistic they begin to exchange energy with
the quintessence field via the coupling term.  At a temperature scale
comparable to the neutrino mass, the neutrinos become
non-relativistic, whilst the quintessence field is dominated by
kinetic energy.  It is at this point that the neutrino mass begins to
evolve significantly.  The details of this behaviour and evolution
depends on the choice of the coupling $\beta$ and the potential
parameter $\sigma$.  In fact, the masses of the neutrinos can be
heavier or lighter in the past depending on the choice of potential
and coupling parameters.

The coupling of neutrinos to dark energy slightly alters the evolution
of the cosmological background.  It was found that similarly to models
with a dark matter/dark energy interaction, the apparent equation of
state measured with Type Ia Supernovae at high redshift can be smaller
than $-1$, without introducing phantom fields, and might even cross
the boundary $w=-1$.

The most obvious modifications to the CMB anisotropy spectrum occur
for large angular scales, with $\ell<100$, where the dominant
contribution to the anisotropies is generated by the Integrated
Sachs-Wolfe Effect (IWS).  This arises due to the evolution of the
gravitational potentials along the photon path from the surface of
last scattering.  The modification to the cosmological background
arising from the neutrino coupling can also have a significant effect
upon the evolution of the perturbations.  We generally observe an
increase in power for $10<\ell<100$, whilst for $\ell<10$ we find
either an excess or reduction in power depending upon our choice of
parameters. For the models where the neutrinos were much heavier in
the past than today, we also observe a slight shift in the peaks and a
modification in their relative amplitude.

The matter power spectrum exhibits free-streaming damping even in the
presence of dark energy--neutrino coupling.  However, since the
damping scale is mainly dependent on the value of the neutrino mass at
the end of their relativistic stage, our results appear similar to the
standard models with CDM and hot dark matter in which the mass is
fixed at the relativistic plateau. It is obvious that the mass infered 
from the damping of the matter power spectrum is, in general, different 
from the neutrino mass measured with experiments in the laboratory. 

We performed a likelihood analysis using SNIa, CMB and LSS
datasets. Initially, we used the standard parameterization for our
cosmological model, characterized by exponential dependence of the
dark energy potential and neutrino mass on the scalar field. For a
flat universe we varied all of the matter parameters, the Hubble
constant, the initial power spectrum spectral index and amplitude and
the instantaneous reionization parameter $z_{\rm re}$. As expected,
the cosmological data did not place strong constraints on our new
coupling parameters.  This is no surprise, since it is well known that
the current best fit analysis of cosmological data can only place an
upper limit on the mass of the neutrino, and a zero neutrino mass is
not excluded by most cosmological data sets. An interesting outcome
was that couplings of order unity are perfectly acceptable with the
actual data.

To proceed, we chose to perform an analysis using two values of the
neutrino mass today, $m_\nu=0.2$~eV and $m_\nu=0.3$~eV, to investigate
whether models of neutrino--dark energy coupling could in principle be
constrained if neutrinos were independently confirmed to have a
significant mass ($m_\nu \gtrsim 0.1$~eV), consistent with current
experiments.

For both the $m_\nu=0.2$~eV and $m_\nu=0.3$~eV models we found that
non-zero values of neutrino coupling strengths of order unity are preferred by the
data.  We also saw that for these models a non--zero value for
$\sigma$ is preferred over the usual cosmological constant, although
$\sigma=0$ is not excluded at the $68\%$ level.  Models with heavier
neutrinos allow stronger constraints to be placed upon the strength of
the coupling.  Indeed, for the $0.3$~eV neutrinos we found that
neutrino--dark energy coupling is preferred at the $1$ sigma
confidence level.

One should note that these constraints rely upon the assumption that
the neutrino mass is known, and that the neutrinos have a mass $m_\nu
\gtrsim 0.1$~eV. Although this assumption is consistent with current
neutrino experiments, we can only make the statement that should the
mass of the neutrino be found to be greater than $0.1$~eV, then
current cosmological data can be used to constrain the strength of any
neutrino--dark energy coupling.



\acknowledgments 
We are grateful to S. Bridle, O. Elgaroy,
H. K. Eriksen, F.K. Hansen, D. Hooper, A. Melchiorri, J. Silk,
C. Skordis and J. Weller for useful discussions. We also thank
A. Lewis for allowing us to use his CAMB quintessence module. AWB is
supported by PPARC.  DFM acknowledge support from the Research Council
of Norway through project number 159637/V30.  DTV acknowledges a
Scatcherd Scholarship.

\newpage
\section*{Erratum} 
\label{sec:erratum}

The original version of this paper contained a typo and a mistake, as noted by  \cite{keum}.  This section contains the corrected equations and Figures, as published in our Erratum \cite{erratum}.

The geodesic equation (23) contains a typo and should read
\begin{equation}
P^0\frac{d P^\rho}{d \tau}+\Gamma^\rho_{\alpha\beta}P^\alpha
P^\beta = -m_\nu^2 \frac{d\ln m_\nu}{d\phi}\frac{\partial \phi}{\partial
x_\rho},
\end{equation}
A subtle error occurred in eq. (24), which should read
\begin{equation}
\frac{dq}{d\tau}=-\frac{1}{2}q\dot{h}_{ij}n_i n_j - a^2 \frac{m^2_\nu}{q}\frac{\partial\ln m_\nu}{\partial \phi}\frac{\partial\phi}{\partial x^i}\frac{\partial x^i}{d\tau}. 
\end{equation}
It follows that a scalar field dependent term should be included in
the Boltzmann equation (26), giving:
\begin{eqnarray}
\frac{\partial \Psi}{\partial \tau} &+& i\frac{q}{\epsilon}\left({\bf
k\cdot n}\right)\Psi \nonumber  + \frac{d\ln f_0}{d\ln q}
\left[\dot\eta-\frac{\dot h+6\dot \eta}{2}\left({\bf k\cdot
n}\right)^2\right]\\ &=&i\frac{q}{\epsilon}\left({\bf k\cdot n}\right)k\frac{a^2 m^2_\nu}{q^2} \frac{\partial \ln m_\nu}{\partial \phi}\frac{d\ln f_0}{d\ln q}\delta\phi
\end{eqnarray}
Therefore, the dipole equation for the neutrino hierarchy derived in
\cite{Ma} is subject to a change represented by a new term once
again dependent on the scalar field:
\begin{equation}
  \dot \Psi_1=\frac{1}{3}\frac{q}{\epsilon}k\left( \Psi_0 - 2\Psi_2  \right)
  -\frac{1}{3}\frac{q}{\epsilon}k\frac{a^2 m^2_\nu}{q^2}\frac{\partial \ln m_\nu}{\partial \phi}\frac{d\ln f_0}{d\ln q}\delta\phi
\end{equation}

\begin{figure}[b!]
\centerline{\scalebox{0.5}{\includegraphics{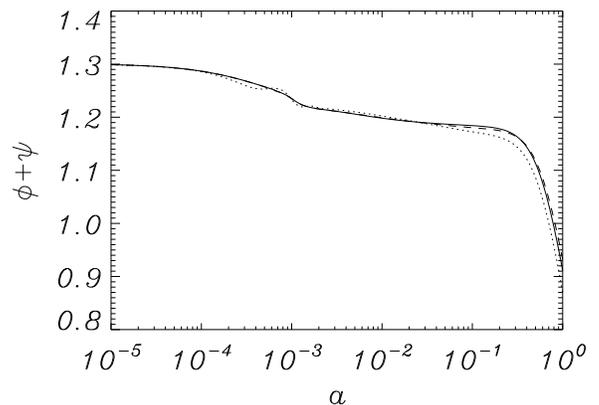}}}
\caption{Evolution of the metric variables $\Phi+\Psi$. Solid line: $\beta=0$, $\lambda = 1$;
short--dashed line: $\beta = 1$, $\lambda=1$; dotted line: $\beta=-0.70$, $\lambda = 1$; 
long--dashed line: $\beta=1$, $\lambda=0.5$. The scale is $k=10^{-3}$~Mpc$^{-1}$.}
\label{fig:fixphipsi}
\end{figure}

This modification will have an effect on the ISW effect, which is less
pronounced than reported in our paper.  The corrected evolution of the
metric variables $\Phi+\Psi$ is shown in Figure \ref{fig:fixphipsi}.

\begin{figure}[htb!]
\centerline{\scalebox{0.5}{\includegraphics{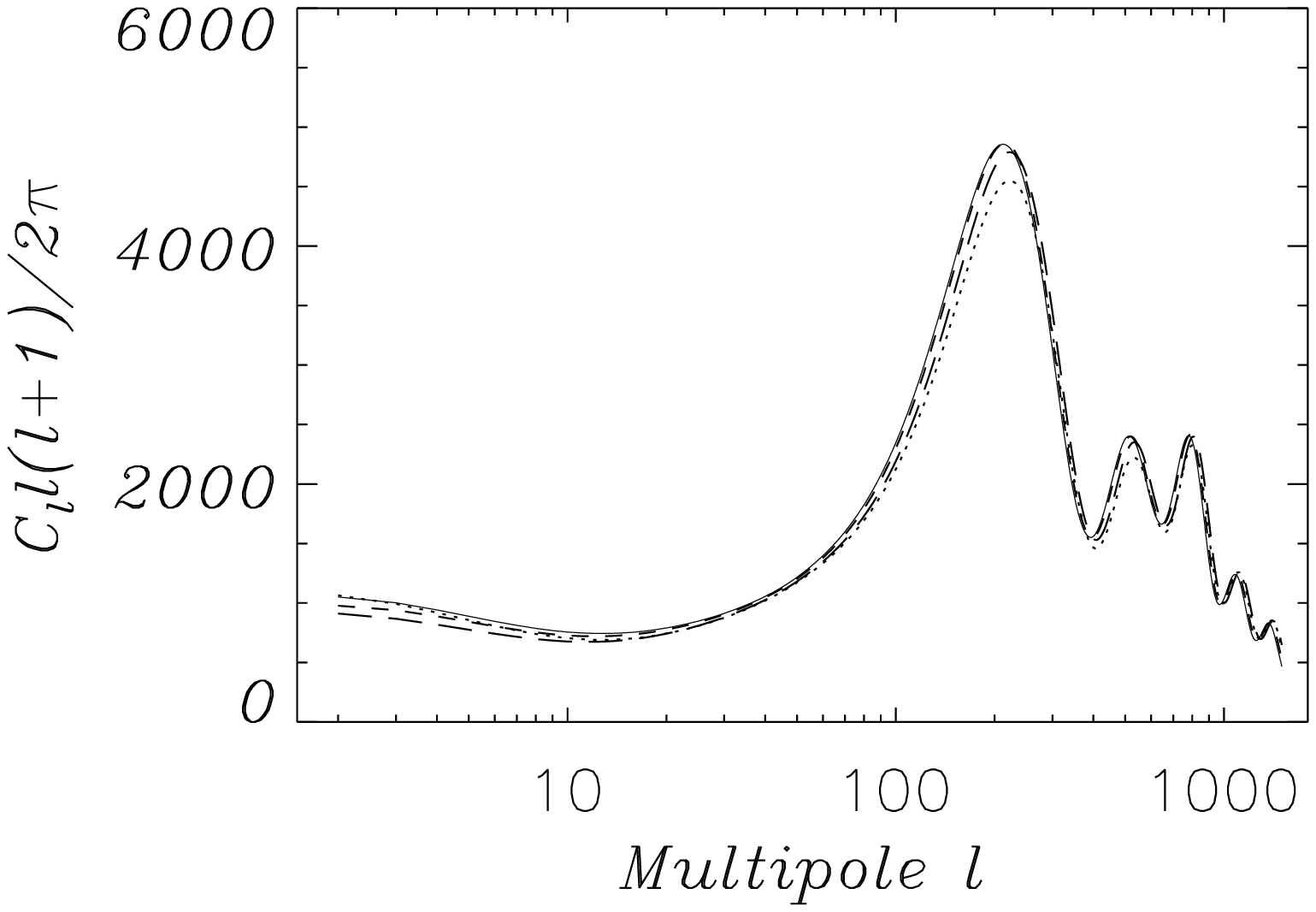}}}
\caption{The CMB anisotropy spectrum (unnormalized) for exponential
coupling and potential. Solid line: $\beta=0$, $\lambda = 1$;
short--dashed line: $\beta = 1$, $\lambda=1$; dotted line:
$\beta=-0.79$, $\lambda = 1$; long--dashed line: $\beta=1$,
$\lambda=0.5$.  }
\label{fig:clsfixed}
\end{figure}

This will effect the anisotropies in the CMB, which are shown in Figure \ref{fig:clsfixed}.

On the other hand, we do not find changes to the matter power spectrum
or the evolution of the neutrino density contrast at the scale given
in Fig. 7 in our original paper. At smaller scales we register small
differences in the neutrino and scalar field density contrasts, which
leads to the mentioned changes in the ISW effect.  Note that the
background evolution reported in \cite{us,usprd} is not affected.

We do not attempt to redo the comparison of our model with data, as
the new WMAP 3-year data have been published since \cite{wmap3}, and
these models where analysed in \cite{keum}.

We are grateful to K. Ichicki for correspondence. The correct
equations (2)-(4) have been derived in \cite{keum}.

\end{document}